\documentclass[10pt]{article}
\usepackage{amsmath}
\usepackage{amssymb}  
\usepackage[T1]{fontenc}
\usepackage{pifont}
\usepackage{pstricks}
\usepackage{amsfonts}
\usepackage{graphicx}
\usepackage{amsmath}
\usepackage{authblk}
\usepackage{pstricks} 
\usepackage{pstricks-add}
\usepackage{caption}
\usepackage{placeins}
\usepackage{pstricks}
\usepackage{pdfpages}
\usepackage{framed}
\usepackage{bm}
\usepackage{epstopdf}
\newtheorem{prop}{Proposition}[section] 
 
\newtheorem{rem}{Remark}[section]

\def\x2dot{\mathop{x}\limits}
\def\y2dot{\mathop{y}\limits}

\def\bfy2dot{\mathop{\bf y}\limits}
\def\z2dot{\mathop{z}\limits}

\def\csi2dot{\mathop{\xi}\limits}
\def\et2dot{\mathop{\eta}\limits}
\def\bet2dot{\mathop{\beta}\limits}
\def\t2dot{\mathop{\theta}\limits}
\def\s2dot{\mathop{\sigma}\limits}
\def\d2dot{\mathop{\delta}\limits}
\def\q2dot{\mathop{q}\limits}
\def\l2dot{\mathop{\lambda}\limits}
\def\ps2dot{\mathop{{\cal E}}\limits}
\def\tet2dot{\mathop{\theta}\limits}
\def\bfx2dot{\mathop{\bf X}\limits}
\def\bfy2dot{\mathop{\bf y}\limits}
\def\bfq2dot{\mathop{\bf q}\limits}
\def\bbfq2dot{\mathop{\bar {\bf q}}\limits}
\def\w2{\mathop{W}\limits}
\def\xgrande2dot{\mathop{\bf X}\limits}
\def\p02dot{\mathop{P}\limits}
\def\a2dot{\mathop{A}\limits}

\title{Nonlinear nonholonomic constraints}
\author{Federico Talamucci}
\affil{DIMAI, Dipartimento di Matematica e Informatica ``Ulisse Dini'',\\
Universit\`a degli Studi di Firenze, Italy\\
tel.~+39 055 2751432, fax +39 055 2751452
\\
e-mail: federico.talamucci@unifi.it}
\date{}

\begin{document}
\bibliographystyle{plain}

\setcounter{equation}{0}
\setcounter{ese}{0}
\setcounter{eserc}{0}
\setcounter{teo}{0}               
\setcounter{corol}{0}
\setcounter{propr}{0}

\maketitle

\vspace{.5truecm}

%Lagrange's equations / nonholonomic dynamical systems nonholonomic systems

\vspace{.5truecm}


\vspace{.5truecm}

\noindent
{\bf Abstract}. 
One of the founders of the mechanics of nonoholonomic systems is Voronec who published in 1901 a significant generalization  
of the ${\check {\rm C}}$aplygin's equations, by removing some restrictive assumptions. In the frame of nonholonomic systems,  the Voronec equations are probably less frequent and common with respect to the prevalent methods of quasi--coordinates (Hamel--Boltzmann equations) and  of the acceleration energy (Gibbs--Appell equations). 
In this paper we start from the case of linear nonholonomic constraints, in order to extend the Voronec equations to nonlinear nonholonomic systems. The comparison between two ways of expressing the equations of motion is performed.
We finally comment that the adopted procedure is appropriated to implement further extensions.

%%%%%%%%%%%%%%%%%%%%%%%%%%%%%%%%%%%%%%%%%%%%%%%%%%%%%%%%%%%%%%%%%%%%%%%%%%%%%%%%%%%%%%%%%%

\section{Introduction}

\noindent
Let us consider $N$ material points $(P_1, m_1)$, $\dots$, $(P_N, m_N)$ whose coordinates in a three-dimensional vector space are listed in the vector ${\bf X}\in {\Bbb R}^{3N}$. 

\noindent
Assume that the system undergoes certain positional constraints, possibly depending on time (i.~e.~fixed or moving constraints involving the coordinates ${\bf X}$): the configurations of the system can be expressed by the representative vector ${\bf X}(q_1,\dots, q_n,t)$, $n\leq 3N$, where $(q_1,\dots, q_n)\in {\cal Q}\subseteq {\Bbb R}^n$ is the set of the local lagrangian coordinates and $t$ appears only if at least one of the geometrical constraints 
depends explicitly on time.
	
\noindent
The Newton's equations $m_i{\p02dot^{..}}_i={\mathcal F}_i+{\mathcal R}_i$, $i=1,\dots, N$, where ${\mathcal F}_i$ and ${\mathcal R}_i$ are respectively the active force and the constraint force concerning $P_i$ can be summarized in ${\Bbb R}^{3N}$ by
\begin{equation}
\label{ne}
{\dot {\mathcal Q}}={\mathcal F}+{\mathcal R}
\end{equation}
where ${\mathcal Q}=(m_1{\dot P}_1, \dots, m_N{\dot P}_N)$ is the representative vector of the linear momentum, ${\mathcal F}=({\mathcal F}_1, \dots, {\mathcal F}_N)$ and ${\mathcal R}=({\mathcal R}_1, \dots, {\mathcal R}_N)$.
As it is known, the space of the all possible velocities consistent with the constraints is in each point the linear space generated by the $n$ vectors $\dfrac{\partial {\bf X}}{\partial q_i}$, $i=1,\dots, n$ and the scalar products of (\ref{ne}) with them lead to the well known equations 
\begin{equation}
\label{eqlagrol}
\dfrac{d}{dt}\dfrac{\partial T}{\partial {\dot q}_i}-\dfrac{\partial T}{\partial q_i}={\mathcal F}^{({\bf q}_i)}
+{\mathcal R}^{({\bf q}_i)} \qquad i=1,\dots,n
\end{equation}
 where $T$ is the kynetic energy $T(q_1,\dots, q_n, {\dot q}_1,\dots, q_n,t)=
\frac{1}{2}{\mathcal Q}\cdot {\dot {\bf X}}$ and for each $i=1,\dots, n$:
\begin{itemize}
	\item[$(a)$]  $\dfrac{d}{dt}\dfrac{\partial T}{\partial {\dot q}_i}-\dfrac{\partial T}{\partial q_i}={\dot {\mathcal Q}}\cdot \dfrac{\partial {\bf X}}{\partial q_i}$,
	\item[$(b)$] ${\mathcal F}^{(q_i)}={\mathcal F}\cdot \dfrac{\partial {\bf X}}{\partial q_i}$,  ${\mathcal R}^{(q_i)}={\mathcal R}\cdot \dfrac{\partial {\bf X}}{\partial q_i}$ are the $i$--th lagrangian component of the active forces - possibly related to a potential scalar function ${\mathcal U}(q_1,\dots, q_n,t)$ - and of the constraint forces, respectively.
\end{itemize}	

\noindent
The constraints are said to be ideal if they play merely the role of restricting the configurations of the system, without entering the possible movements of it: as it is well known, in the holonomic case this is equivalent to the vanishing of  
the lagrangian components of ${\mathcal R}$, since the set of the possible displacements 
corresponds to the linear space ${\mathcal T}_{\bf X}$ (the tangent space) generated by $\frac{\partial {\bf X}}{\partial q_1}$, $\dots$,
$\frac{\partial {\bf X}}{\partial q_n}$.
In that case, the $n$ equations (\ref{eqlagrol}) contain 
precisely the $n$ unknown quantities $q_1$, $\dots$, $q_n$.	
A simple and suitable way to move forward more general systems consists in keeping in mind points $(a)$ and $(b)$ listed above: to identify the set of displacements (vectors) along which the constraints forces are said to be ideal - by reasonable motivations -  and to develop the calculus of (\ref{ne}) along those directions, finding the additional terms which generally appear, besides the Lagrangian binomial. First of all, such a way to proceed will make us retrieve the linear nonholonomic Voronec equations, as we explain hereafter.
	
\noindent
On the basis of that frame we let
additional constraints of kinematical type to be present: such a situation can be formulated by operating directly on the lagrangian coordinates and by adding the equations
\begin{equation}
\label{constr}
\Phi_j(q_1,\dots, q_n, {\dot q}_1, \dots, {\dot q}_n)=0, \quad j =1,\dots, k
\end{equation}
where $k<n$ and ${\dot {\bf q}}=({\dot q}_1, \dots, {\dot q}_n)\in {\Bbb R}^\ell$ are the generalized velocities. 
The given functions $\Phi_j$ are assumed to be independent w.~r.~t.~the kinetic variables, namely the jacobian matrix $J_{({\dot q}_1,\dots, {\dot q}_n)}(\Phi_1, \dots, \Phi_m)$ attains its maximum rank $m$. 

\noindent
Without limiting the generality of our discussion, we can assume that the regularity condition of the nonholonomic constraints holds for the $k\times k$ submatrix formed by the last $k$ columns, so that (\ref{constr}) can be written in the form 
\begin{equation}
\label{constrexpl}
\begin{cases}
{\dot q}_{m+1}=\alpha_1(q_1, \dots, q_n, {\dot q}_1, \dots, {\dot q}_m) \\
\dots  \\
{\dot q}_{m+k}=\alpha_k(q_1, \dots, q_n, {\dot q}_1, \dots, {\dot q}_m)
\end{cases}
\end{equation}
with $m=n-k$ somehow decreases the range of freedom of the system, when the kinematical constraints (\ref{constr}) are encompassed, 

\noindent
Our main aim is to formulate the equations  of motion if  the nonholonomic conditions (\ref{constr}) are embraced in the dynamics. It is evident that the main question to be rivised in (\ref{eqlagrol}) concerns with the new set of possible velocities, owing to (\ref{constr}) (or (\ref{constrexpl})): the right idea to pursue is the same as in the holonomic case, of null constraint forces along all the possible displacement.

\noindent
Actually, joining (\ref{eqlagrol}) with (\ref{constr}) produces a set of $n+m$ equations, with the $2n$ unknows quantities $q_1(t)$, $\dots$, $q_n(t)$ and ${\mathcal R}^{(q_1)}$, $\dots$,   ${\mathcal R}^{(q_n)}$ which cannot be claimed to be null and which are difficult to model at this stage of the problem.

\noindent
Although the most common technique in nonholonomic problems makes use of the so called quasi--coordinates and quasi--velocities (see, among others, \cite{gant}, \cite{green}, \cite{lurie}, \cite{neimark}, \cite{pars}), the point of view adopted here is to employ only the lagrangian coordinates and velocities already present in the equations.
More precisely, the implicit conditions (\ref{constr}) correspond to make the set of $m$ velocities ( $({\dot q}_1, \dots, {\dot q}_m)$ in the selected case) as the independent ones and assuming all values in ${\Bbb R}^m$, whereas the remaining velocities $({\dot q}_{m+1}, \dots, {\dot q}_{m+k=n})$ have to assume the values (\ref{constrexpl}), in each position $(q_1,\dots, q_n)$, at any time $t$ and for each set $({\dot q}_1, \dots, {\dot q}_m)$, in order to be consistent with the kinematical constraints.
 
\noindent
This point of view dates back to Voronec, whose equations will be reproduced in the next Section. They deals with a simpler case with respect to our assumptions: the rest of the paper is devoted to make a generalization of Voronec's equations. Unexpectedly, they are not as common as other classical sets for nonholonomic systems (Maggi \cite{maggi}, Hamel \cite{hamel}, Appell \cite{appell},...) and probably the most known version of them is the specific case introduced by ${\check {\rm C}}$aplygin, which will be mentioned hereafter.

 \section{The Voronec equations for linear kinematical constraints}
 
 \noindent
 The equations derived by Voronec in \cite{voronec} concerns the case of linear nonholonomic constraints:
 
\begin{equation}
\begin{cases}
\label{constrexpllin}
{\dot q}_{m+1}=\sum\limits_{i=1}^m\alpha_{1,i}(q_1,\dots, q_n){\dot q}_i, \\
\dots \\
{\dot q}_{m+k}=\sum\limits_{i=1}^m\alpha_{k,i}(q_1,\dots, q_n){\dot q}_i \\
\end{cases}
\end{equation}
(we use again the greek letter $\alpha$, where the double subscript distinguishes the linear case).
The lagrangian expression of the velocity of the system is 
$$
{\dot {\bf X}}={\widehat {\dot {\bf X}}}+\frac{\partial {\bf X}}{\partial t}
$$
where ${\widehat {\dot {\bf X}}}=\sum\limits_{i=1}^n  {\dot q}_i \dfrac{\partial {\bf X}}{\partial q_i}$ is any velocity 
consistent with the instantaneous configuration of the system (i.~e.~at a blocked time $t$) and the second term appears in case of mobile constraints. Owing to (\ref{constrexpllin}), one has
\begin{equation}
\label{subspace}
{\widehat {\dot {\bf X}}}=
\sum\limits_{i=1}^m  {\dot q}_i\left(
\dfrac{\partial}{\partial q_i}{\bf X}(q_1,\dots, q_n,t)+
\sum\limits_{j=1}^{k}\alpha_{j,i}(q_1,\dots, q_n)\dfrac{\partial}{\partial q_{m+j}}{\bf X}(q_1,\dots, q_n,t)\right)
\end{equation}
The arbitrariness of $({\dot q}_1, \dots, {\dot q}_m)$ makes the set of possible displacements, at any blocked time $t$, the linear subspace of ${\mathcal T}_{\bf X}$ generated by the $m$ vectors of (\ref{subspace})  (for $i=1,\dots,m$) in brackets. 

\noindent
We require the constraint forces to play the same ideal role as described in the Introduction: the natural extension of ideal constraint demands that the constraint forces have no lagrangian components on that subspace, that is 
\begin{equation}
\label{vincanolid}
{\mathcal R}\cdot \left(
\dfrac{\partial {\bf X}}{\partial q_i}+
\sum\limits_{j=1}^{k}\alpha_{j,i}\dfrac{\partial {\bf X}}{\partial q_{m+j}}\right)=
{\mathcal R}^{(q_i)}+\sum\limits_{j=1}^{k}\alpha_{j,i}{\mathcal R}^{(q_{m+j})}=0, \qquad i=1,\dots, m.
\end{equation} 
Equivalently, the constraint forces are ideal if the lagrangian components 
${\mathcal R}^{({\bf q})}=\left( {\mathcal R}^{(q_1)}, \dots, {\mathcal R}^{(q_n)}\right)^T$ are linear combinations of the rows of the matrix $\left( A \; |\, - {\Bbb I}_k \right)$ or, that is the same, they verify
\begin{equation}
\label{idealmatr}
\left( {\Bbb I}_m \;\vert \; A\right){\mathcal R}^{({\bf q})}={\bf 0}
\end{equation}
where $A$ is the $k\times m$ matrix of elements $\alpha_{j,i}$ appearing in (\ref{constrexpllin}) and ${\Bbb I}_r$ is the unit matrix of order $r$, $r=k,m$.

\noindent
From the point of view of the energy of the system, we see that (\ref{vincanolid}) entails $\sum\limits_{i=1}^n{\mathcal R}^{(q_i)}{\dot q}_i=0$, so that the energy is not dissipated.

\begin{rem}
%Since ${\dot {\bf X}}=\sum\limits_{i=1}^n {\dot q}_i\dfrac{\partial {\bf X}}{\partial q_i}+\dfrac{\partial {\bf X}}{\partial t}$, 
An alternative way to attain the same set (\ref{subspace}) is to consider the vectors
\begin{equation}
\label{subspace1}
\sum\limits_{i=1}^m  {\dot q}_i \dfrac{\partial {\dot {\bf X}}}{\partial {\dot q}_i}
\end{equation}
for arbitrary $({\dot q}_1, \dots, {\dot q}_m)$ and taking account of (\ref{constrexpllin}).
The linearity of the velocity of the system with respect to the generalized velocities ${\dot q}_i$ is a well--known property in the lagrangian formalism.
Such an evidence will be convenient in order to access the more general case discussed in the following.

\end{rem}
At this point, it makes sense to multiply the Newton's equation (\ref{ne}) by the $m$ vectors in brackets in (\ref{subspace}), for each $i$, in order to achieve the following equations of motion:
\begin{equation}
\label{voronec0}
\dfrac{d}{dt}\dfrac{\partial T}{\partial {\dot q}_i}-\dfrac{\partial T}{\partial q_i}+
\sum\limits_{j=1}^k \alpha_{j,i}\left(
\dfrac{d}{dt}\dfrac{\partial T}{\partial {\dot q}_{m+j}}-\dfrac{\partial T}{\partial q_{m+j}}\right)=
{\mathcal F}^{(q_i)}+\sum\limits_{j=1}^k \alpha_{j,i} {\mathcal F}^{(q_{m+j})}, \qquad i=1,\dots, m.
\end{equation}
The $m$ equations (\ref{voronec0}) joined with (\ref{constrexpllin}) consist of a set of $m+k=n$ differential equations in the $n$ unknown quantities $q_1(t)$, $\dots$, $q_n(t)$.

\noindent
One advantage of (\ref{voronec0}) is the direct appearance of coefficients $\alpha_{j,i}$ in the equations of motions, 
owing to the explicit form of the coinstraints conditions (\ref{constrexpllin}). If the coinstraints are expressed by implicit linear conditions like $\sum\limits_{j=1}^n a_{i,j}{\dot q}_j=0$, $i=1,\dots, m$, it is necessary to calculate the vectors orthogonal to the rows of the matrix $a_{i,j}$ in order to get  the coefficients entering the equations of motion (such a  procedure must be adopted, as an instance, for writing the Maggi's equations).

\begin{rem}
In (\ref{voronec0}) one can recover the case when one of the contraints (\ref{constrexpllin}), say the $r$--th one, is integrable, that is it  exists a function $f^{(r)}$ such that $\alpha_{r,i}(q_1,\dots, q_m)=\frac{\partial f^{(r)}}{\partial q_i}$ for any $i=1,\dots, m$. The term in (\ref{voronec0}) for $j=r$ is 
$\frac{\partial f^{(r)}}{\partial q_i}(\frac{d}{dt}\frac{\partial T}{\partial {\dot q}_r}-\frac{\partial T}{\partial q_r})$ 
and this corresponds, as it is known, to add to the system ${\bf X}(q_1,\dots, q_n,t)$ the holonomic constraint $q_r=f^{(r)}(q_1,\dots, q_m)$.
\end{rem}

\noindent
Following \cite{neimark}, equations {\ref{voronec0}} can be formulated in terms of the reduced function
\begin{equation}
\label{tridlin}
T^*(q_1,\dots, q_n, {\dot q}_1, \dots, {\dot q}_m, t)=T(q_1, \dots, q_n, {\dot q}_1, \dots, {\dot q}_m, {\dot q}_{m+1}(\cdot), \dots, {\dot q}_{m+k}(\cdot), t)
\end{equation}
where ${\dot q}_{m+j}(\cdot)$, $j=1,\dots, k$, stands for ${\dot q}_{m+j}(q_1,\dots, q_n, {\dot q}_1, \dots, {\dot q}_m)$,  in accordance with (\ref{constrexpl}). Simple calculations lead to the Voronec's equations of motion 

\begin{equation}
\label{voronec}
\dfrac{d}{dt}\dfrac{\partial T^*}{\partial {\dot q}_i}-\dfrac{\partial T^*}{\partial q_i}
-\sum\limits_{\nu=1}^k\alpha_{\nu,i}\dfrac{\partial T^*}{\partial q_{m+\nu}}
-\sum\limits_{\nu=1}^k\sum\limits_{j=1}^m 
\beta_{ij}^\nu{\dot q}_j\dfrac{\partial T}{\partial {\dot q}_{m+\nu}}=
{\mathcal F}^{(q_i)}+\sum\limits_{j=1}^{k} \alpha_{j,i} {\cal F}^{(q_{m+j})}
\quad i=1,\dots, m
\end{equation}
where
\begin{equation}
\label{beta}
\beta_{ij}^\nu(q_1,\dots, q_n)=
\dfrac{\partial \alpha_{\nu, i}}{\partial q_j}-
\dfrac{\partial \alpha_{\nu,j}}{\partial q_i}+
\sum\limits_{\mu=1}^k\left(
\dfrac{\partial \alpha_{\nu, i}}{\partial q_{m+\mu}}\alpha_{\mu,j}-
\dfrac{\partial \alpha_{\nu, j}}{\partial q_{m+\mu}}\alpha_{\mu,i}
\right).
\end{equation}
In the terms $\dfrac{\partial T}{\partial {\dot q}_{m+\nu}}$ the arguments ${\dot q}_{m+1}$, $\dots$, ${\dot q}_n$ have to be removed by taking advantage of (\ref{constrexpllin}).

\begin{rem}
Under the same assumptions of Remark 2.1, it is worth checking the effect of an integrable constraint in (\ref{voronec}): by defining the reduced function 
$$
T_r=T^*(q_1,\dots, q_m,\dots, q_{r-1}, f^{(r)}(q_1,\dots, q_m), q_{r+1}, \dots, q_n, {\dot q}_1, \dots, {\dot q}_m,t)
$$
which does not contain $q_r$,  one has for $\nu=r$:
$$
\dfrac{d}{dt}\dfrac{\partial T^*}{\partial {\dot q}_i}-\dfrac{\partial T^*}{\partial q_i}	-\alpha_{r,i}\dfrac{\partial T^*}{\partial q_{m+r}}=
\dfrac{d}{dt}\dfrac{\partial T_r}{\partial {\dot q}_i}-\dfrac{\partial T_r}{\partial q_i}
$$ and $\beta_{ij}^r=0$. 
Clearly, if any of the constraints (\ref{constrexpllin}) is integrable, the left side of (\ref{voronec}) reduces to the lagrangian binomial  $\frac{d}{dt}\frac{\partial {\hat T}}{\partial {\dot q}_i}-\frac{\partial {\hat T}}{\partial q_i}$ characteristic of holonomic systems, where 
$$
{\hat T}(q_1, \dots, q_m, {\dot q}_1, \dots, {\dot q}_m, t)=T^*(q_1, \dots, q_m, f^{(1)}(q_1, \dots, q_m), \dots, f^{(k)}(q_1, \dots, q_m), {\dot q}_1, \dots, {\dot q}_m, t).
$$ 
\end{rem}

\noindent
Although the equations of motion (\ref{voronec}) do not contain the velocities ${\dot q}_{m+1}$, $\dots$, $\dot q_{m+k}$, 
system (\ref{voronec}) is still coupled with the constraints expressions (\ref{constrexpllin}), because of the presence of $q_{m+1}$, $\dots$, $q_{m+k}$. A special case, formulated by  ${\check {\rm C}}$aplygin some years before the Voronec equations, consists in assuming that the lagrangian coordinates $q_1$, $\dots$, $q_m$ do not occur in $T$, in the coefficients $\alpha_{j,i}$, $j=1, \dots, m$ nor in the forces ${\cal F}^{(q_i)}$ for any $i=1,\dots, m$.
In that case, (\ref{voronec0}) reduces to the ${\check {\rm C}}$aplygin equations \cite{capligin}

\begin{equation}
\label{capligin}
\dfrac{d}{dt}\dfrac{\partial T^*}{\partial {\dot q}_i}-\dfrac{\partial T^*}{\partial q_i}
-\sum\limits_{\nu=1}^k\sum\limits_{j=1}^m 
\left(\dfrac{\partial \alpha_{\nu, i}}{\partial q_j}-
\dfrac{\partial \alpha_{\nu,j}}{\partial q_i}\right)
{\dot q}_j\dfrac{\partial T}{\partial {\dot q}_{m+\nu}}=
{\cal F}^{(q_i)}+\sum\limits_{j=1}^{k} \alpha_{j,i} {\cal F}^{(q_{m+j})}
\quad i=1,\dots, m
\end{equation}
where $T^*=T^*(q_1,\dots, q_m, {\dot q}_1, \dots, {\dot q}_m,t)$, the same for the forces terms.
The clear advantage is that the set (\ref{capligin}) is now disentangled from (\ref{constrexpllin}) 
and one needs to solve only $m$ differential equations in order to solve the motion.

\noindent
Assumptions for (\ref{capligin}) may appear demanding: 
it is worthwhile to remark that ${\check {\rm C}}$aplygin systems are not so uncommon in real examples, if the lagrangian coordinates are properly chosen.

\noindent
Lastly, let us comment how is it framed (\ref{voronec}) within a more general situation, where (\ref{constrexpllin}) are replaced by the linear implicit conditions $\sum\limits_{i=1}^m\Phi_{j,i}(q_1,\dots, q_n){\dot q}_i=0$, $j=1,\dots, k$. Using a vector--matrix notation for simplicity, we write ${\mit \Phi}({\bf q}){\dot {\bf q}}={\bf 0}$, where ${\mit \Phi}$ is the $k\times n$ matrix ${\mit \Phi}$ of elements $(\Phi_{j,i})$ and ${\bf q}=(q_1,\dots, q_n)$. If the rank of ${\mit \Phi}$ is maximal, $m$ independent vectors $(\gamma_{1,j}, \dots, \gamma_{n,j})$, $j=1,\dots, m$, orthogonal to the rows of ${\mit \Phi}$ can be found, so that  ${\mit \Phi}{\mit \Gamma}={\Bbb O}$, with $\Gamma$ $n\times m$ matrix of elements $\gamma_{i,j}$  and ${\Bbb O}_{k,m}$ zero matrix $k\times m$. The set (\ref{subspace}) is replaced by $\sum\limits_{i=1}^m {\dot q}_i \sum\limits_{j=1}^n\gamma_{j,i}\frac{\partial {\bf X}}{\partial q_i}$ and the ideal constraints assumption (\ref{vincanolid}) by $\sum\limits_{j=1}^n\gamma_{j,i}{\cal R}^{(q_j)}=0$, $i=1,\dots, m$, so that the $m$ equations of motion are
\begin{equation}
\label{eqvett}
{\mit \Gamma}^T \left( \dfrac{d}{dt}\nabla_{\dot {\bf q}}T-\nabla_{\bf q}T-{\cal F}^{(\bf q)}\right)={\bf 0}
\end{equation}
with obvious meaning of symbols. As we said before, the special case leading to (\ref{voronec}) does not require the calculation fo ${\mit \Gamma}$: indeed, it is ${\mit \Phi}=\left( A \; |\, - {\Bbb I}_k \right)$ and ${\mit \Gamma}^T=\left( {\Bbb I}_m \;\vert \; A\right)$, where $A$ is the $k\times m$ matrix of elements $\alpha_{j,i}$ appearing in (\ref{constrexpllin}) and ${\Bbb I}_r$ is the unit matrix of order $r$, $r=k,m$ (see also (\ref{idealmatr})).
Incidentally, we remark that if an invertible change of coordinates ${\bar q}_i={\bar q}_i(q_1,\dots, q_n)$, $i=1,\dots, n$ is implemented, then the constraints equations move to ${\bar {\mit \Phi}}({\bar {\bf q}}){\dot {\bar {\bf q}}}={\bf 0}$, with ${\bar {\mit \Phi}}={\mit \Phi}
({\bf q}({\bar {\bf q}}))(J_{\bar {\bf q}}{\bf q})$ (the latter is the jacobian matrix of entries $\frac{\partial q_i}{\partial {\bar q}_j}$, $i,j=1,\dots, n$). At the same time, the new matrix of orthogonal vectors is ${\bar {\mit \Gamma}}=(J_{\bf q}{\bar {\bf q}}){\mit \Gamma}$: it follows that the writing of the equations of motion in terms of ${\bar {\bf q}}$, that is 
$$
{\bar {\mit \Gamma}}^T \left( \frac{d}{dt}\nabla_{\dot {\bar {\bf q}}}T-\nabla_{\bar {\bf q}}T
-{\cal F}^{({\bar {\bf q}})}\right)={\mit \Gamma}^T(J_{\bf q}{\bar {\bf q}})^T (J_{\bar {\bf q}}{\bf q})^T
\left( \frac{d}{dt}\nabla_{\dot {\bf q}}T-\nabla_{\bf q}T-{\cal F}^{(\bf q)}\right)
$$
exhibits a balance between the change of ${\mit \Gamma}$ (showing the inverse of the jacobian matrix of the transformation
${\bar {\bf q}}({\bf q})$) and the change of the lagrangian equations, in the same way as the jacobian matrix. This means an invariant behaviour of equations (\ref{eqvett}), which can be easily explained in terms of lagrangian components and contravariant components of (\ref{ne}) in the basis $\dfrac{\partial {\bf X}}{\partial q_i}$, $i=1,\dots, n$. 

\begin{rem}
Concerning (\ref{constrexpllin}), it is clear that 
a general transformation ${\bar {\bf q}}({\bf q})$ does not preserve the explicit arrangement of the constraints and the simplified path to (\ref{voronec0}): if this is demanded, only partial changes of coordinates ${\bar q}_i(q_1,\dots, q_m)$, $i=1,\dots, m$, ${\bar q}_j=q_j$, $j=m+1,\dots, n$, can be considered.
\end{rem}

\section{The nonlinear case}

\noindent
The aim is to extend equation of type (\ref{voronec}) to the case of nonlinear constraints (\ref{constrexpl}). We refer to \cite{papastr} for a comprehensive list of references abuot nonlinear nonholonomic systems, where the scarcity of literature - especially in english - on such an important topic is underlined. 
The starting point comes from Remark 2.1: the constraints forces are going to be tested with the set of vectors (\ref{subspace1}), which takes the form
\begin{equation}
\label{subspace2}
{\widehat {\dot {\bf X}}}=
\sum\limits_{i=1}^m  {\dot q}_i\left(
\dfrac{\partial}{\partial q_i}{\bf X}(q_1,\dots, q_n,t)+
\sum\limits_{j=1}^{k}\dfrac{\partial \alpha_j}{\partial {\dot q}_i}(q_1,\dots, q_n,  {\dot q}_1, \dots, {\dot q}_m)\dfrac{\partial}{\partial q_{m+j}}{\bf X}(q_1,\dots, q_n,t)\right)
\end{equation}
For arbitrary  $m$--uples $({\dot q}_1, \dots, {\dot q}_m)$ the set (\ref{subspace2}) plays the role of the totality of the possible velocities consistent with the constraints at each position $(q_1, \dots, q_n)$ and, when the configuration space is freezed at a time $t$. As in the linear case, assumption (\ref{idealnonlin}) entails ${\cal R}\cdot {\hat {\dot {\bf X}}}=0$, that is the power of the constraint forces is zero, with respect to all the possible displacements consistent to any blocked configuration of the system.

\noindent
We are motivated to state that the constraint forces ${\cal R}$ are ideal if they are orthogonal to $m$ the vectors in brackets in (\ref{subspace2}): 
\begin{equation}
\label{idealnonlin}
{\cal R}\cdot 
\left(
\dfrac{\partial {\bf X}}{\partial q_i}+
\sum\limits_{j=1}^{k}\dfrac{\partial \alpha_j}{\partial {\dot q}_i}\dfrac{\partial {\bf X}}{\partial q_{m+j}}
\right)=0, \qquad \textrm{for each}\;\;i=1,\dots, m.
\end{equation}
In an equivalent way, we can refer to the lagrangian components of the constraint forces extending condition (\ref{idealmatr}) to nonlinear constraints, by replacing the matrix $A$ with the $k\times m$ matrix of entries $\dfrac{\partial \alpha_i}{\partial {\dot q}_j}$, $i=1,\dots,k$, $j=1,\dots, m$. In this way, 
the generalized constraint forces ${\cal R}^{({\bf q})}$ are orthogonal to the $m$ vectors $(\underbrace{0,\dots, \overbrace{1}^{j-th}, \dots, 0}_m,\frac{\partial \alpha_1}{\partial {\dot q}_j}, \dots, 
\frac{\partial \alpha_k}{\partial {\dot q_j}})$, $j=1,\dots, m$, whereas they are linear combinations of the $k$ vectors $(\frac{\partial \alpha_i}{\partial {\dot q}_1}, \dots, 
\frac{\partial \alpha_i}{\partial {\dot q_m}}\underbrace{0,\dots, \overbrace{-1}^{i-th}, \dots, 0}_k)$, $i=1,\dots, k$.

\begin{prop}
The equations of motion of the system ${\bf X}({\bf q},t)$ subject to the nonlinear kinematical constraints (\ref{constrexpl}) are 

\begin{equation}
\label{vnl}
\dfrac{d}{dt}\dfrac{\partial T^*}{\partial {\dot q}_i}-\dfrac{\partial T^*}{\partial q_i}
-\sum\limits_{\nu=1}^k\dfrac{\partial T^*}{\partial q_{m+\nu}}\dfrac{\partial \alpha_\nu}{\partial {\dot q_i}}
-\sum\limits_{\nu=1}^k \dfrac{\partial T}{\partial {\dot q}_{m+\nu}} B_{i}^\nu
={\cal F}^{(q_i)}+\sum\limits_{j=1}^k \dfrac{\partial \alpha_j}{\partial {\dot q}_i} {\cal F}^{(q_{m+j})}, 
\qquad i=1,\dots, m
\end{equation}
where
\begin{equation}
\label{b}
B_{i}^{\nu}(q_1,\dots, q_n, {\dot q}_1, \dots, {\dot q}_m)=\sum\limits_{r=1}^m
\left( \dfrac{\partial^2 \alpha_\nu}{\partial {\dot q}_i \partial q_r}{\dot q}_r +
\dfrac{\partial^2 \alpha_\nu}{\partial {\dot q}_i \partial {\dot q}_r}{\q2dot^{..}}_r\right)
-\dfrac{\partial \alpha_\nu}{\partial q_i}+
\sum\limits_{\sigma=1}^k\left( \dfrac{\partial^2 \alpha_\nu}{\partial {\dot q}_i \partial q_{m+\sigma}}\alpha_\sigma-
\dfrac{\partial \alpha_\sigma}{\partial {\dot q}_i}\dfrac{\partial \alpha_\nu}{\partial q_{m+\sigma}}\right)
\end{equation}
and $T^*$ is the reduced function
\begin{equation}
\label{trid}
T^*(q_1,\dots, q_n, {\dot q}_1, \dots, {\dot q}_m, t)=T(q_1, \dots, q_n, {\dot q}_1, \dots, {\dot q}_m, \alpha_1(\cdot), \dots, \alpha_k(\cdot), t)
\end{equation}
intending for each $\alpha_j(\cdot)$, $j=1,\dots, k$, the dependence on $(q_1,\dots, q_n, {\dot q}_1, \dots, {\dot q}_m)$ according to (\ref{constrexpl}). 
\end{prop}

\noindent
{\bf Proof}. 
In the same way as we did for (\ref{voronec0}), we multiply (\ref{ne}) by the $m$ vectors appearing in (\ref{subspace2}), for each $i=1,\dots, m$, in order to get, reminding  (\ref{idealnonlin}):

\begin{equation}
\label{voronec0nonlin}
\dfrac{d}{dt}\dfrac{\partial T}{\partial {\dot q}_i}-\dfrac{\partial T}{\partial q_i}+
\sum\limits_{j=1}^k \dfrac{\partial \alpha_j}{\partial {\dot q}_i}\left(
\dfrac{d}{dt}\dfrac{\partial T}{\partial {\dot q}_{m+j}}-\dfrac{\partial T}{\partial q_{m+j}}\right)=
{\cal F}^{(q_i)}+\sum\limits_{j=1}^k \dfrac{\partial \alpha_j}{\partial {\dot q}_i} {\cal F}^{(q_{m+j})}, 
\qquad i=1,\dots, m
\end{equation}
The following relations

\begin{equation}
\label{r12}
\begin{array}{ll}
\dfrac{\partial T}{\partial {\dot q}_i}=\dfrac{\partial T^*}{\partial {\dot q}_i}-\sum\limits_{\nu=1}^k 
\dfrac{\partial T}{\partial {\dot q}_{m+\nu}}\dfrac{\partial \alpha_\nu}{\partial {\dot q}_i}, & i=1,\dots, m \\
\\
\dfrac{\partial T}{\partial q_i}=\dfrac{\partial T^*}{\partial q_i}-\sum\limits_{\nu=1}^k 
\dfrac{\partial T}{\partial {\dot q}_{m+\nu}}\dfrac{\partial \alpha_\nu}{\partial q_i}, & i=1,\dots, n
\end{array}
\end{equation}
allow us to write (\ref{voronec0nonlin}) in terms of $T^*$: in order to get (\ref{vnl}) it suffices to have in mind that
$$
\begin{array}{l}
-\dfrac{d}{dt}\left(\dfrac{\partial T}{\partial {\dot q}_{m+\nu}}\dfrac{\partial \alpha_\nu}{\partial {\dot q}_i}\right)+
\dfrac{d}{dt}\left(\dfrac{\partial T}{\partial {\dot q}_{m+\nu}}\right)\dfrac{\partial \alpha_\nu}{\partial {\dot q}_i}=
-\dfrac{\partial T}{\partial {\dot q}_{m+\nu}}
\dfrac{d}{dt}\left(\dfrac{\partial \alpha_\nu}{\partial {\dot q}_i}\right)=\\
\\=
-\dfrac{\partial T}{\partial {\dot q}_{m+\nu}}\left(
\sum\limits_{s=1}^m
\left( 
\dfrac{\partial^2 \alpha_\nu}{\partial {\dot q}_i \partial q_s}{\dot q}_s +
\dfrac{\partial^2 \alpha_\nu}{\partial {\dot q}_i \partial {\dot q}_s}{\q2dot^{..}}_s
\right)
+
\sum\limits_{j=1}^k
\dfrac{\partial^2 \alpha_\nu}{\partial {\dot q}_i \partial q_{m+j}}\alpha_j
\right). \qquad \square
\end{array}
$$
Whenever $\alpha_\nu$ are the linear functions $\sum_{j=1}^m\alpha_{\nu, j}{\dot q}_j$ of (\ref{constrexpllin}) for any $\nu=1,\dots, k$, then (\ref{vnl}) and (\ref{b}) reproduce exactly (\ref{voronec}) and (\ref{beta}): we have in that case, for any $i=1,\dots, m$ and $\nu = 1,\dots, k$
$$
\begin{array}{l}
\dfrac{\partial \alpha_\nu}{\partial {\dot q}_i}=\alpha_{\nu,i}, \qquad
\sum\limits_{s=1}^m\dfrac{\partial^2 \alpha_\nu}{\partial {\dot q}_i \partial q_s}{\dot q}_s-
\dfrac{\partial \alpha_\nu}{\partial q_i}
=\sum\limits_{j=1}^m \left(
\dfrac{\partial \alpha_{\nu,i}}{\partial q_j}-\dfrac{\partial \alpha_{\nu, j}}{\partial q_i}\right){\dot q}_j, \qquad
\dfrac{\partial^2 \alpha_\nu}{\partial {\dot q}_i \partial {\dot q}_s}=0,\\
\\
\sum\limits_{j=1}^k\left( \dfrac{\partial^2 \alpha_\nu}{\partial {\dot q}_i \partial q_{m+j}}\alpha_j-
\dfrac{\partial \alpha_j}{\partial {\dot q}_i}\dfrac{\partial \alpha_\nu}{\partial q_{m+j}}\right)=
\sum\limits_{\mu=1}^k \sum\limits_{j=1}^m
\left(\dfrac{\partial \alpha_{\nu, i}}{\partial q_{m+\mu}}\alpha_{\mu,j}-
\dfrac{\partial \alpha_{\nu, j}}{\partial q_{m+\mu}}\alpha_{\mu,i}\right){\dot q}_j.
\end{array}
$$

\noindent
The $m$ equations (\ref{vnl}), containing the $n$ unknown functions $q_1(t), \dots, q_n(t)$, have to be coupled with (\ref{constrexpl}) in order to solve the problem. Even in this case, if the functions appearing in the equations exhibit  special sets of variables, a reduction of the problem can be made analogously to (\ref{capligin})  and the 
${\check {\rm C}}$aplygin's equations can be formally extended to the nonlinear case.
Apart from this, the main point we are interested in is the mathematical problem arisen from (\ref{vnl}), specifying our analysis on the case where $T$ does not contain explicitly $t$.

\subsection{The case of fixed constraints}

\noindent
The presence of the second derivatives ${\q2dot^{..}}_s$ in the coefficients $B_i^\nu$ seems unusual and may somehow alter the structure we usually meet in ordinary lagrangian equations, even in the linear nonholonomic case, where the second derivatives of the unknown functions originate only from the first term in (\ref{voronec}). Clearly, the existence and uniqueness of the solution is closely related to the way the second derivatives appear in the equations.
Hence, we do not find pointless to derive the equations of motion following a different way, in the special (but significant)
case of fixed holonomic constraints, namely ${\bf X}={\bf X}(q_1, \dots, q_n)$.

\noindent
Focussing on the left side of (\ref{ne}), let us make use of the $3N$ vector ${\bf X}^{(m)}=(m_1P_1,\dots, m_N P_N)$ (position vector equipped by masses of the points), so that
\begin{eqnarray}
\nonumber
{\cal Q}=
{\dot {\bf X}^{(m)}}&=&\sum\limits_{i=1}^m\dfrac{\partial {\bf X}^{(m)}}{\partial q_i}{\dot q}_i+\sum\limits_{j=1}^k
\dfrac{\partial {\bf X}^{(m)}}{\partial q_{m+j}}\alpha_j, \\
\nonumber
{\dot {\cal Q}}={\bfx2dot^{..}}^{(m)}&=&
\sum\limits_{i=1}^m \dfrac{\partial {\bf X}^{(m)}}{\partial q_i}{\q2dot^{..}}_i
+\sum\limits_{i,j=1}^m \dfrac{\partial^2 {\bf X}^{(m)}}{\partial q_i \partial q_j}{\dot q}_i {\dot q}_j 
+2\sum\limits_{i=1}^m \sum\limits_{j=1}^{k}\dfrac{\partial^2 {\bf X}^{(m)}}{\partial q_i \partial q_{m+j}}{\dot q}_i\alpha_j
+\sum\limits_{i,j=1}^k \dfrac{\partial^2 {\bf X}^{(m)}}{\partial q_{m+i} \partial q_{m+j}}\alpha_i \alpha_j+\\
\label{calqdot}
&+&\sum\limits_{j=1}^k\dfrac{\partial {\bf X}^{(m)}}{\partial q_{m+j}}\left(
\sum\limits_{i=1}^m \left(\dfrac{\partial \alpha_j}{\partial q_i}{\dot q}_i+
\dfrac{\partial \alpha_j}{\partial {\dot q}_i} {\q2dot^{..}}_i\right)+ 
\sum\limits_{\nu=1}^k \dfrac{\partial \alpha_j}{\partial q_{m+\nu}}\alpha_\nu\right)
\end{eqnarray}

\begin{prop}
Define for each $i,j,k=1,\dots, n$
	\begin{equation}
	\label{gxi}
	g_{i,j}(q_1,\dots, q_n)=\dfrac{\partial {\bf X}^{(m)}}{\partial q_i}\cdot \dfrac{\partial {\bf X}}{\partial q_j}, \quad 
	\xi_{i,j,k}(q_1, \dots, q_n)=\dfrac{\partial^2{\bf X}^{(m)}}{\partial q_i\partial q_j}\cdot \dfrac{\partial {\bf X}}{\partial q_k}. 
	\end{equation}	
Then,the equations of motions of the system ${\bf X}({\bf q})$ subject to the nonlinear kinematical constraints (\ref{constrexpl}) can be written in the following form:
\begin{equation}
\label{vnl2}
\sum\limits_{\nu=1}^m \left( C_i^\nu {\q2dot^{..}}_\nu+
\sum\limits_{\mu=1}^m D_i^{\nu,\mu} {\dot q}_\nu{\dot q}_\mu + E_i^\nu {\dot q}_\nu\right) +G_i={\cal F}^{(q_i)}+\sum\limits_{j=1}^k \dfrac{\partial \alpha_j}{\partial {\dot q}_i} {\cal F}^{(q_{m+j})}, 
\qquad i=1,\dots, m
\end{equation}
where the coefficients, depending on $(q_1,\dots, q_n, {\dot q}_1, \dots, {\dot q}_m)$, are defined as 
$$
\begin{array}{l}
C_i^\nu=\sum\limits_{r,s=1}^k\left(g_{i, \nu}+
g_{i,m+r}\dfrac{\partial \alpha_r}{\partial {\dot q}_\nu}
+g_{m+r,\nu}\dfrac{\partial \alpha_r}{\partial {\dot q}_i}+ g_{m+s,m+r}\dfrac{\partial \alpha_s}{\partial {\dot q}_i}\dfrac{\partial \alpha_r}{\partial {\dot q}_\nu}\right), \\
D_i^{\nu,\mu}=\xi_{\nu,\mu,i}+\sum\limits_{r=1}^k
\xi_{\nu,\mu,m+r}\dfrac{\partial \alpha_r}{\partial {\dot q}_i}, \\
E_i^\nu =  \sum\limits_{r,s=1}^k \left( 2\left(
\xi_{\nu,m+r,i}+\xi_{\nu,m+r,m+s} \dfrac{\partial \alpha_s}{\partial {\dot q}_i} \right) \alpha_r
+\left( g_{m+r,i}+ g_{m+r,m+s}\dfrac{\partial \alpha_s}{\partial {\dot q}_i}\right)\dfrac{\partial \alpha_r}{\partial q_\nu}\right),
\\
G_i=\sum\limits_{r,s,p=1}^k\left(
\left(\xi_{m+r,m+s,i}+ \xi_{m+r, m+s,m+p} \dfrac{\partial \alpha_p}{\partial {\dot q}_i}\right) \alpha_r\alpha_s
+\left( g_{m+r,i}+g_{m+r,m+p}\dfrac{\partial \alpha_p}{\partial {\dot q}_i}\right)
\dfrac{\partial \alpha_r}{\partial q_{m+s}}\alpha_s \right).
\end{array}
$$
\end{prop}

\noindent
{\bf Proof}.
Recalling (\ref{subspace2}) and implementing the products
${\dot {\cal Q}}\cdot
(\frac{\partial {\bf X}}{\partial q_i}+
\sum\limits_{j=1}^{k}\frac{\partial \alpha_j}{\partial {\dot q}_i}\frac{\partial {\bf X}}{\partial q_{m+j}})$ for each $i=1,\dots, m$ where ${\dot {\cal Q}}(q_1,\dots, q_n, {\dot q}_1, \dots, {\dot q}_m, \alpha_1 (\cdot), \dots, \alpha_k(\cdot))$ is the function in (\ref{calqdot}) and $(\cdot)$ stands for $(q_1,\dots, q_n, {\dot q}_1, \dots, {\dot q}_m)$, straightforward calculations easily drive to (\ref{vnl2}). $\quad \square$

\noindent
Equations (\ref{vnl2}), possibly expressed in terms of quasi--velocities, are the same equations that can be deduced from the Gauss principle, as for instance discussed in \cite{benenti}.

\begin{rem}
The procedure of Proposition 3.2 corresponds to the method of Appell: indeed, the energy of accelerations is 
$S=\dfrac{1}{2} {\dot {\cal Q}}\cdot \bfx2dot^{..}$ and $\dfrac{\partial S}{\partial {\q2dot^{..}}_i}=
{\dot {\cal Q}}\cdot \dfrac{\partial \bfx2dot^{..}}{\partial {\q2dot^{..}}_i}={\dot {\cal Q}}\cdot
(\frac{\partial {\bf X}}{\partial q_i}+
\sum\limits_{j=1}^{k}\frac{\partial \alpha_j}{\partial {\dot q}_i}\frac{\partial {\bf X}}{\partial q_{m+j}})$: in other words, the  calculus to implement in order to write the Appell equations is exactly the same illustrated in the proof of Proposition 3.2. 
\end{rem}

\noindent
Equations (\ref{vnl}) and equations (\ref{vnl2}), describing the same motion, show dissimilarities, which can be mainly reported to how the constraint equations $\alpha_j$ appear.
Thus, it is not pointless to develop explicitly the calculations in (\ref{vnl}), in order to check if redundant terms are present. In terms of the defined quantities and in the examined case, the kinetic energy is
$T=\dfrac{1}{2}\sum\limits_{i,j}^n g_{i,j}{\dot q}_i {\dot q}_j$ and (\ref{trid}) takes the form 
\begin{equation}
\label{tfixed}
T^*=\dfrac{1}{2}\sum\limits_{i,j}^m g_{i,j}{\dot q}_i {\dot q}_j+
\sum\limits_{\nu=1}^k \alpha_\nu \left( 
\dfrac{1}{2}\sum\limits_{\mu=1}^k g_{m+\nu,m+\nu}\alpha_\mu +\sum\limits_{j=1}^m g_{j,m+\nu}{\dot q}_j \right)
\end{equation}

\begin{prop}
Assume that equations (\ref{vnl}) are written with $T^*$ as in (\ref{tfixed}). Then, for each $i=1,\dots, m$ the terms $-\sum\limits_{\nu=1}^k \dfrac{\partial T}{\partial {\dot q}_{m+\nu}} B_{i}^\nu$ correspond to the sum of the terms
$$
\begin{array}{ll}
(i) &-\sum\limits_{r,s=1}^m \sum\limits_{\nu,\mu, \sigma=1}^k \left(g_{r,m+\nu}{\dot q}_r+g_{m+\nu, m+\mu} \alpha_\mu\right)
\left( \dfrac{\partial^2 \alpha_\nu}{\partial {\dot q}_i \partial q_s}{\dot q}_s +
\dfrac{\partial^2 \alpha_\nu}{\partial {\dot q}_i \partial {\dot q}_s}{\q2dot^{..}}_s+
\dfrac{\partial^2 \alpha_\nu}{\partial {\dot q}_i \partial q_{m+\sigma}}\alpha_\sigma\right),\\ %%%%% B C D E
(ii) & \sum\limits_{r,s=1}^m \sum\limits_{\nu,\mu, \sigma=1}^k \left(g_{r,m+\nu}{\dot q}_r+g_{m+\nu, m+\mu} \alpha_\mu\right)
\dfrac{\partial \alpha_\sigma}{\partial {\dot q}_i}\dfrac{\partial \alpha_\nu}{\partial q_{m+\sigma}} \\ %%%%%%% F
(iii) & \sum\limits_{r=1}^m\sum\limits_{\nu=1}^k \left( g_{r,m+\nu}{\dot q}_r+g_{m+\nu,m+\mu}\alpha_\mu\right)\dfrac{\partial \alpha_\nu}{\partial q_i} %%%%% A 
\end{array}
$$
which appear with opposite sign in $\dfrac{d}{dt}\left(\dfrac{\partial T^*}{\partial {\dot q}_i}\right)$ {\rm(}terms $(i)${\rm)}, 
in $-\dfrac{\partial T^*}{\partial q_i}$ {\rm(}terms $(ii)${\rm)} and in
$-\sum\limits_{\nu=1}^k\dfrac{\partial T^*}{\partial q_{m+\nu}}\dfrac{\partial \alpha_\nu}{\partial {\dot q_i}}$ {\rm(}terms $(iii)${\rm)} respectively. Therefore, all the terms of $-\sum\limits_{\nu=1}^k \dfrac{\partial T}{\partial {\dot q}_{m+\nu}} B_{i}^\nu$ vanish and the remaining terms of (\ref{vnl}) coincide precisely with (\ref{vnl2}).
\end{prop}

\noindent
{\bf Proof.} The check is a not short but simple calculus based on the following preliminary formulae (whenever (\ref{tfixed})is assumed)
$$
\begin{array}{l}
\dfrac{\partial T^*}{\partial {\dot q}_i}=
\sum\limits_{r=1}^m \sum\limits_{\nu,\mu=1}^k
\left(
\left( g_{r,i} + g_{r,m+\nu}\dfrac{\partial \alpha_\nu}{\partial {\dot q_i}}\right){\dot q}_i
+ \left( g_{m+\nu, m+\mu} \dfrac{\partial \alpha_\nu}{\partial {\dot q}_i}+g_{i,\mu} \right) \alpha_\mu \right)
\quad i=1,\dots, m \\
\\
\dfrac{\partial T^*}{\partial q_i}=
\sum\limits_{r,s=1}^m  \sum\limits_{\nu,\mu=1}^k \left(
\dfrac{1}{2}\left(
\dfrac{\partial g_{r,s}}{\partial q_i}{\dot q}_r{\dot q}_s+
\dfrac{\partial g_{m+\nu,m+\mu}}{\partial q_i}\alpha_\nu \alpha_\mu \right)
+\left( g_{m+\nu, m+\mu}\alpha_\mu + g_{r,m+\nu}{\dot q}_r\right)\dfrac{\partial \alpha_\nu}{\partial q_i}
+\dfrac{\partial g_{r,m+\nu}}{\partial q_i}{\dot q}_r \alpha_\nu \right) \\
\quad i=1,\dots, n \\
\left.\dfrac{\partial T}{{\dot q}_{m+\nu}}
\right\vert_{{\dot q}_{m+\mu}=\alpha_\mu, \mu=1,\dots, k}=
\sum\limits_{r=1}^mg_{m+\nu, r}{\dot q}_r +\sum\limits_{\mu=1}^k g_{m+\nu, m+\mu}\alpha_\mu
\quad \nu=1,\dots, k.
\end{array}
$$
Once the just written expressions have been placed in (\ref{vnl}), the terms declared in the statement of the Proposition cancel and the remaining terms match with (\ref{vnl2}). $\quad \square$

\section{Conclusions and feasible generalizations}

\noindent
The framework depicted by (\ref{constrexpl}) is quite general, since the constraints equations are assumed to be independent.
The Voronec equations develop such a starting point, without employing quasi--coordinates and quasi--velocities.

\noindent
The classic Voronec equations for linear kinematical constraints have been extended to the nonlinear case by (\ref{vnl}). At the same time, in a special case (fixed constraints) the comparison with the set of equations depicting the same motion and 
present in literature in an apparently dissimilar form made us conclude that several terms in (\ref{vnl}) are redundant. 
However, this is strictly connected to the specific selection summarized by (\ref{trid}). The two different ways of drawing the equations of motions reflect the two points of view of keeping the lagrangian structure of the equations save for additional terms (equations (\ref{vnl})), or basing directly on the D'Alembert's principle  he one and 

\noindent
Two main points can be planned in order to examine the question in a more general frame: 
\begin{enumerate}
	\item the holonomic constraints depend explicitly on time $t$,
	\item even the nonholonomic constraints depend on time.
\end{enumerate}

\noindent
The first topic is actually already sketched by equations (\ref{vnl}), where the manifold of configurations is allowed to be mobile. Nevertheless, the expression of the kinetic energy is no longer (\ref{tfixed}) and $0$--degree and $1$--degree terms with respect to ${\dot q}_1, \dots, {\dot q}_m$ have to be added. Hence, Propositions 3.2 and 3.3 need to be rearranged in order to check the possible deletions of terms.

\noindent
As regards the second issue, whenever one (or more) of (\ref{constrexpl}) is replaced by ${\dot q}_{m+j}=\alpha_j(q_1, \dots, q_n, {\dot q}_1, \dots, {\dot q}_m, t)$, the set of velocities consistent with the instantaneous configuration of the system is still (\ref{subspace}), but the equations of motion cannot be longer written in the form (\ref{vnl}). More precisely, relations (\ref{r12}) still hold, but not the successive formula.

\noindent
A final interesting theme which can be treated in a natural way by means of the introduced approach concerns higher order constraints equations, of the type ${\mit \Phi}_j
\left(q_1, \dots, q_n, {\dot q}_1, \dots, {\dot q}_n, \dots, 
\dfrac{d^p q_1}{dt^p}, \dots, \dfrac{d^p q_n}{dt^p}, t \right)=0$, $p\geq 2$: the key point is to extend (\ref{subspace1}) to higher derivatives, by virtue of the lagrangian known property 
$\dfrac{\partial \bfx2dot^{\kappa \cdot}}{\partial {\q2dot^{\kappa \cdot}}_i}=\dfrac{\partial {\bf X}}{\partial q_i}$, where $\kappa \cdot$ stands for the sequence of $\kappa$ dots, $\kappa=1,2,\dots$.

\end{document}